%Paper: hep-th/9411181
%From: Vitaly Tarasov <tarasov@tftxb.helsinki.fi>
%Date: Thu, 24 Nov 1994 18:08:32 +0200

%%%%%%%%%%%%%%%%%%%%%%%%%%%%%%%%%%%%%%%%%%%%%%%%%%%%%%%%%%%%%%%%%%%%%%%%%
%									%
%  Solutions to the Quantized Knizhnik-Zamolodchikov			%
%  Equation and the Bethe Ansatz					%
%									%
%  by V.Tarasov	 and  A.Varchenko					%
%									%
%  to appear in Proceedings of XX-th ICGTMP (Osaka, July 4-9, 1994)	%
%  									%
%  (amstex.tex 2.1, amsppt.sty 2.1 are required)			%
%									%
%%%%%%%%%%%%%%%%%%%%%%%%%%%%%%%%%%%%%%%%%%%%%%%%%%%%%%%%%%%%%%%%%%%%%%%%%

\mag 1200

\input amstex
\input amsppt.sty

\overfullrule 0pt

\NoRunningHeads
\TagsOnRight

\hsize 5in
\vsize 7.08333in

\expandafter\ifx\csname osaka.def\endcsname\relax \else\endinput\fi
\expandafter\edef\csname osaka.def\endcsname{%
 \catcode`\noexpand\@=\the\catcode`\@\space}
\catcode`\@=11

\mathsurround 1.6pt
\font\bbf=cmbx12  \font\foliorm=cmr9

\def\hcor#1{\advance\hoffset by #1}
\def\vcor#1{\advance\voffset by #1}
\let\bls\baselineskip \let\dwd\displaywidth \let\ignore\ignorespaces
\def\vsk#1>{\vskip#1\bls} \let\adv\advance 
\def\vv#1>{\vadjust{\vsk#1>}\ignore} \def\vvv#1>{\vadjust{\vskip#1}\ignore}
\def\vvn#1>{\vadjust{\nobreak\vsk#1>\nobreak}\ignore}
\def\vvvn#1>{\vadjust{\nobreak\vskip#1\nobreak}\ignore}
\def\setnormalbls{\edef\normalbls{\bls\the\bls}}
\def\setmaths{\edef\maths{\mathsurround\the\mathsurround}}

\let\vp\vphantom  \let\^\negthickspace
\let\nl\newline \let\nt\noindent 
\def\nn#1>{\noalign{\vskip #1pt}} \def\NN#1>{\openup#1pt}
 
\let\Sum\sum \def\sum{\Sum\limits} 
\let\Prod\prod \def\prod{\Prod\limits} \let\Int\int \def\int{\Int\limits}
\def\tsum{\mathop{\tsize\Sum}\limits} 

\let\=\m@th \def\&{.\kern.1em} \def\>{\!\;} \def\:{\!\!\;}
\def\~{\leavevmode\raise.16ex\hbox{\=${-}$}}
\def\^{\leavevmode\kern.04em\raise.16ex\hbox{--}}

\def\center{\begingroup\leftskip 0pt plus \hsize \rightskip\leftskip
 \parindent 0pt \parfillskip 0pt \def\\{\break}}
\def\endcenter{\endgraf\endgroup}

\ifx\plainfootnote\undefined \let\plainfootnote\footnote \fi
\expandafter\ifx\csname amsppt.sty\endcsname\relax
 
\else \fi

\newbox\dib@x
\def\hleft#1:#2{\setbox\dib@x\hbox{$\dsize #1\quad$}\rlap{$\dsize #2$}
 \kern-2\wd\dib@x\kern\dwd}
\def\hright#1:#2{\setbox\dib@x\hbox{$\dsize #1\quad$}\kern-\wd\dib@x
 \kern\dwd\kern-\wd\dib@x\llap{$\dsize #2$}}

\newbox\sectb@x
\def\sect#1 #2\par{\removelastskip\vskip.8\bls
 \vtop{\bf\setbox\sectb@x\hbox{#1} \parindent\wd\sectb@x
 \ifdim\parindent>0pt\adv\parindent.5em\fi\item{#1}#2\strut}%
 \nointerlineskip\nobreak\vtop{\strut}\nobreak\vskip-.6\bls\nobreak}

\newbox\t@stb@x
\def\gadv{\global\advance} \def\gad#1{\gadv#1 1} 
\def\l@b@l#1#2{\def\n@@{\csname #2no\endcsname}%
 \if *#1\gad\n@@ \expandafter\xdef\csname @#1@\endcsname{\the\n@@}%
 \else\expandafter\ifx\csname @#1@\endcsname\relax\gad\n@@
 \expandafter\xdef\csname @#1@\endcsname{\the\n@@}\fi\fi}
\def\l@bel#1#2{\l@b@l{#1}{#2}\[#1]}
\def\[#1]{\csname @#1@\endcsname}
\def\(#1){{\setbox\t@stb@x\hbox{\[#1]}\ifnum\wd\t@stb@x=0\rm({\bf???})\else
 \rm(\[#1])\fi}}
\def\dff{\expandafter\d@f} \def\d@f{\expandafter\def}
\def\edff{\expandafter\ed@f} \def\ed@f{\expandafter\edef}

\newcount\Sno \newcount\Lno \newcount\Fno
\def\Sect{\gad\Sno\sect{\the\Sno.} }
\def\l@F#1{\l@bel{#1}F} \def\<#1>{\l@b@l{#1}F>}
\def\Tag#1{\tag\l@F{#1}} \def\Tagg#1{\tag"\llap{\rm(\l@F{#1})}"}
\def\Th#1{Theorem \l@bel{#1}L} \def\Lm#1{Lemma \l@bel{#1}L}
 \def\Cj{Conjecture}

\def\Par{\par\medskip} \def\setparindent{\edef\Parindent{\the\parindent}}
\def\Appendix{\Sno=64\let\p@r@\z@ %\parindent
 \def\Sect{\gad\Sno\Fno=0\Lno=0 \sect{}\hskip\p@r@ Appendix \char\the\Sno}
 \def\l@b@l##1##2{\def\n@@{\csname ##2no\endcsname}%
 \if *##1\gad\n@@
 \expandafter\xdef\csname @##1@\endcsname{\char\the\Sno.\the\n@@}%
 \else\expandafter\ifx\csname @##1@\endcsname\relax\gad\n@@
 \expandafter\xdef\csname @##1@\endcsname{\char\the\Sno.\the\n@@}\fi\fi}}

\newcount\Rno
\def\Ci#1{\l@bel{#1@Ref}R}
\def\Cite#1{{\=$^{\,\l@bel{#1@Ref}R}$}}
\def\myRefs{\=\tenpoint\sect{} % \hskip\Parindent
 References\par
 \def\k@yf@##1{\hss##1.\enspace}
 \def\widest##1{\setbox\t@stb@x\hbox{\tenpoint\k@yf@{##1}}%
  \refindentwd\wd\t@stb@x}
 \let\keyformat\k@yf@
 \def\Key##1{\setbox\t@stb@x\hbox{\[##1@Ref]}\ifnum\wd\t@stb@x=0
  \no{\bf???}\else \no\[##1@Ref]\fi}}
\let\alb\allowbreak

\let\o\circ \let\x\times \let\ox\otimes
 
 \let\ge\geqslant
\let\der\partial  \let\8\infty
\let\bra\langle \let\ket\rangle
 
 \let\map\mapsto 
 
\let\=\m@th  \def\_#1{_{\rlap{$\ssize#1$}}}

\def\lsym#1{#1\alb\ldots#1\alb}
\def\lc{\lsym,}  \def\lx{\lsym\x} \def\lox{\lsym\ox}
 
\def\E(#1){\mathop{\hbox{\rm End}\,}(#1)} 
 \def\const{\hbox{\rm const}}

\def\1{^{-1}} \def\vst#1{{\lower2.1pt\hbox{$\bigr|_{#1}$}}}

\let\gm\gamma \let\Gm\Gamma
 \let\Dl\Delta
 \let\eps\varepsilon \let\epsilon\eps

\let\la\lambda \let\La\Lambda
  
\let\pho\phi \let\phi\varphi

\def\C{\Bbb C}
\def\R{\Bbb R}

\def\Z{\Bbb Z}

\def\Rb{\text{\rm R}} 
\def\S{\bold S}

\def\g{\frak g}
\def\CC{\frak C}

\def\CCm{\frak C(z,\mu)}
\def\tb{\bar t}
\def\thu{\hat\tau}

\def\d{^{\raise.5ex\hbox{\=$\sssize\dagger$}}}
\def\D{^{\raise.5ex\hbox{\=$\sssize\ddagger$}}}
\def\Ka{K\d}

\def\Vl{V_\la}
\def\Rv{R^{\vp1}}

\def\Cl{\C^{\,\ell}}
\def\CN{\C^{N+1}}
\def\ZN{\Z^N_{\ge 0}}

\def\dt{\,d^\ell t}
\def\ts{t^{\sssize\star}}
\def\zs{z^{\sssize\star}}
\def\sing{\hbox{\rm Sing\>}V}
\def\singl{\hbox{\rm Sing\>}\Vl}
\def\dims{\dim\,\singl}

\def\egv/{eigenvector}
\def\eva/{eigenvalue}
\def\eq/{equation}
\def\lhs/{the left hand side}
\def\rhs/{the right hand side}
\def\Rm/{{\=$R$-matrix}}
\def\Rms/{{\=$R$-matrices}}
\def\conv/{convenient}
\def\sol/{solution}
\def\as/{asymptotic}
\def\asol/{\as/ \sol/}
\def\rep/{representation}
\def\ir/{irreducible}
\def\YB/{Yang-Baxter \eq/}
\def\sym/{symmetric}
\def\hom/{homomorphism}
\def\aut/{automorphism}
\def\gb/{generated by}
\def\wrt/{with respect to}
\def\perm/{permutation}
\def\fn/{function}
\def\var/{variable}
\def\resp/{respectively}
\def\pol/{polynomial}
\def\tri/{trigonometric}
\def\rat/{rational}
\def\reg/{regular}
\def\prop/{proportional}
\def\inrp/{integral \rep/}
\def\st/{such that}
\def\corr/{correspond}
\def\lex/{lexicographical}
\def\ndg/{nondegenerate}
\def\neib/{neighbourhood}
\def\raf/{\rat/ \fn/}

\def\difl/{differential}
\def\dif/{difference}
\def\deq/{\dif/ \eq/}
\def\dsc/{discrete}
\def\cc/{compatibility condition}
\def\fd/{finite-dimensional}
\def\gv/{generating vector}
\def\wt/{weight}
\def\m/{module}
\def\hw/{highest \wt/}
\def\hwm/{\hw/ \gm/}
\def\wtd/{\wt/ decomposition}
\def\phf/{phase \fn/}
\def\cp/{critical point}
\def\ncp/{\ndg/ \cp/}
\def\Bv/{Bethe vector}
\def\Ba/{Bethe ansatz}
\def\Bae/{\Ba/ \eq/}
\def\asex/{\as/ expansion}
\def\msd/{method of steepest descend}
\def\off/{offdiagonal}
\def\Vval/{{\=$V\!\!\;$-valued}}

\def\tens/{tensor product}
\def\gm/{{\,\=$\g$-\m/}}
\def\Ym/{{\=$Y\!$-\m/}}

\def\KZ/{{\sl KZ\/}}
\def\qKZ/{{\sl qKZ\/}}
\def\KZo/{\qKZ/ operator}
\def\KZv/{Knizh\-nik-Zamo\-lod\-chi\-kov}

\def\gg{\frak{gl}_2}
\def\gl{\frak{gl}_{N+1}}

\def\tram/{transfer-matrix}
\def\qsc/{quantum spin chain model}

\def\TFT/{Research Insitute for Theoretical Physics}
\def\HY/{University of Helsinki}
\def\AoF/{the Academy of Finland}
\def\myaddress/{P.O\&Box 9 (Siltavuorenpenger 20\,\,C), SF\^00014, \HY/,
 Finland}
\def\myemail/{tarasov\@phcu.helsinki.f{i}}
\def\SPb/{St\&Petersburg}
\def\home/{Physics Department, \SPb/ University, \SPb/ \,198904, Russia}
\def\UNC/{Department of Mathematics, University of North Carolina}
\def\ChH/{Chapel Hill}
\def\avaddress/{\ChH/, NC 27599, USA}
\def\grant/{NSF Grant DMS\^9203929}

\def\Tar/{V\:\&Tarasov}
\def\Fadd/{L\&D\&Faddeev}
\def\Smir/{F\:\&A\&Smirnov}
\def\Fre/{I\&Frenkel}
\def\Resh/{N\&Reshetikhin}
\def\Reshy/{N\&\:Yu\&Reshetikhin}
\def\Takh/{L\&A\&Takhtajan}
\def\Kir/{A\&N\&Kirillov}
\def\Varch/{A\&\:Varchenko}
\def\Varn/{A\&N\&\:Varchenko}

\def\CMP/{Commun. Math. Phys.}

\let\foliofont@\foliorm
\let\logo@\relax
\let\m@k@h@@d\makeheadline \let\m@k@f@@t\makefootline
\def\makeheadline{\ifnum\pageno=1\headline={\hfil}\fi\m@k@h@@d}
\def\makefootline{\ifnum\pageno=1\footline={\hfil}\fi\m@k@f@@t}

\setnormalbls \setmaths \setparindent
\csname osaka.def\endcsname

\def\narrower{\leftskip.083333\hsize \rightskip\leftskip}

\document

\center
{\bbf
Solutions to the Quantized Knizhnik-Zamolodchikov
\vsk.25>
Equation and the Bethe-Ansatz}
\vsk1.5>\vfill
{\smc \Tar/}
\vsk.5>
{\it \TFT/\\ \myaddress/
\vsk.3>
\home/}
\vsk.7>
{\eightpoint and}
\vsk.7>
{\smc \Varch/}
\vsk.5>
{\it \UNC/\\ \avaddress/}
\endcenter
\vsk1.5>\vfill
{\eightpoint
\centerline{\smc Abstract}
\vsk.6>

{\narrower\nt
We give an \inrp/ for \sol/s to the quantized \KZv/ \eq/ (\qKZ/) associated
with the Lie algebra $\gl$. Asymptotic \sol/s to \qKZ/ are
constructed. The leading term of an \asol/ is the \Bv/ -- an \egv/ of
the \tram/ of a \qsc/. We show that the norm of the \Bv/ is
equal to the product of the Hessian of a suitable \fn/ and an explicitly
written \raf/. This formula is a generalization of the Gaudin-Korepin
formula for a norm of the \Bv/. We show that, generically, the \Bv/s form
a base for the $\gg$ case.
\vsk>\vfill}}
\vsk>

\sect{} Introduction
\par\nt
The quantized \KZv/ \eq/ (\qKZ/) is a holonomic system of \deq/s
introduced recently in$^{\,\Ci{FR},\,\Ci{S},\,\Ci{JM}}\!$.
\qKZ/ inherits many remarkable properties of the \difl/ \KZv/ \eq/ (\KZ/).
In particular, there is an \inrp/ for \sol/s to \qKZ/\,\Cite{TV} associated to
$\gl$ or $U_q(\gl)$ which has a similar structure of an integrand as an \inrp/
for \sol/s to \KZ/$^{\,\Ci{SV},\,\Ci{V0}}\!$. The $N=1$ case was considered
earlier in$^{\,\Ci{M},\,\Ci{R2},\,\Ci{V1}}\!$.
(Cf\. also$^{\,\Ci{S2},\,\Ci{JMKQ},\,\Ci{KQ}}\!$).
Asymptotic \sol/s to \qKZ/\,\Cite{TV2} obtained from an \inrp/ establish
a connection  between the \Ba/ and \qKZ/. The $\gl$ and $U_q(\gl)$
analogues of the Gaudin-Korepin formula for the norm of the \Bv/ can be
proved up to a multiplicative constant in this framework.
Similar results for \KZ/ are obtained in$^{\Ci{RV},\,\Ci{V2}}\!$.
\par
In this paper we consider only the case of \qKZ/ associated to $\gl$ although
almost all the results can be lifted to the $U_q(\gl)$ case,
cf\&$^{\Ci{TV},\,\Ci{TV2}}\!$.
Section 2 is based on\,\Cite{TV} and Sections 3,4 are based on\,\Cite{TV2}.

\Sect \qKZ/ associated with $\gl$
\par\nt
Let $\g=\gl$ with the canonical generators $\{E_{ij}\}$.
Let $Y$ be the \corr/ing Yangian with a coproduct $\Dl$. Let
$\phi:Y\to U(\g)$ be the natural \hom/ and let $\theta_z$, $z\in\C$, be
the canonical \aut/ of $Y$.
Set $\phi_z=\phi\o\theta_z$. For any two \hwm/s $V_1,V_2$ with
\gv/s $v_1,v_2$, \resp/, there is a unique \Rm/
$\Rv_{V_1V_2}(z)\in\E(V_1\ox V_2)$, \st/ for any $X\in Y$
$$
\kern-1em\Rv_{V_1V_2}(z_1-z_2)\,(\phi_{z_1}\ox\phi_{z_2})\o\Dl(X)=
(\phi_{z_1}\ox\phi_{z_2})\o\Dl'(X)\,\Rv_{V_1V_2}(z_1-z_2)
\Tag{R1}
$$
in $\E(V_1\ox V_2)$ and $\Rv_{V_1V_2}(z)\,v_1\ox v_2=v_1\ox v_2$.
Here $\Dl'=P\o\Dl$ and $P$ is a \perm/ of factors in $Y\ox Y$.
$\Rv_{V_1V_2}(z)$ preserves the weight decomposition of a \gm/ $V_1\ox V_2$;
its restriction to any weight subspace of $V_1\ox V_2$ is a \raf/ in $z$.
 For any $\mu\in\CN$ introduce
$L(\mu)=\exp\Bigl(\>\sum^{N+1}_{i=1}\mu_iE_{ii}\Bigr)$, which is well defined
in any \hwm/.
\par
Let $V_1\lc V_n$ be \hwm/s, $V=V_1\lox V_n$. Let $\rho_i:\E(V_i)\to\E(V)$ be
embeddings as tensor factors. Set
$R_{ij}(z)=\rho_i\ox\rho_j(\Rv_{V_iV_j}(z))$ and $L_i(\mu)=\rho_i(L(\mu))$.
Let $p\in\C$ and $z=(z_1\lc z_n)$.
The operators
$$
\align
K_i(z;p)=R_{i,i-1}(z_i-z_{i-1}+p)\ldots R_{i1}(z_i-z_1+p)\x{}&\\
{}\x L_i(\mu)\,R\1_{ni}(z_n-z_i) \ldots R\1_{i+1,i}(z_{i+1}-z_i)&
\Tagg{kzo}
\endalign
$$
are called the {\it \KZo/s}. Denote by $Z_i$ the {\=$p\>$-shift} operator:
$$
Z_i : \Psi (z_1\lc z_n) \map \Psi(z_1\lc z_i+p \lc z_n)\,.
$$
The {\it quantized \KZv/ \eq/}\,\Cite{FR} is the holonomic system of
\deq/s for a \Vval/ \fn/ $\Psi(z;p)$:
$$
Z_i\Psi(z;p)=K_i(z;p)\Psi(z;p)\,,\qquad i=1\lc n\,.
\Tag{qkz}
$$

\Sect Integral \rep/s for solutions to \qKZ/
\par\nt
Let $t=(t_1\lc t_\ell)$.
Let $Q_a$ be the {\=$p\>$-shift} operator \wrt/ a variable $t_a$.
Let $\Phi(t,z;p)$ be a meromorphic scalar \fn/ and $w(t,z;p)$
a \Vval/ \raf/ in $t,z$. Say that $\Phi(t,z;p)w(t,z;p)$ gives an {\it \inrp/}
for \sol/s to system \(qkz) if
$$
Z_i(\Phi w)-K_i\>\Phi w = \sum_{a=1}^\ell\,
\bigl(Q_a(\Phi w_{ai})-\Phi w_{ai}\bigr)\,,\qquad i=1\lc n\,,
$$
for suitable \raf/s $w_{ai}(t,z;p)$. $\Phi(t,z;p)$ and $w(t,z;p)$ are
called the {\it \phf/} and the {\it \wt/ \fn/}, \resp/.
\proclaim{\Th{TV}}
There exists an \inrp/ for \sol/s to \qKZ/ \(qkz) associated with $\gl$.
\endproclaim
\nt
We describe the \inrp/ for \sol/s to \qKZ/ more explicitly below.
\Par
 Fix $\la\in\ZN$. Let $\La(1)\lc\La(n)\in\CN$ be \hw/s of \gm/s
$V_1\lc V_n$, \resp/. Let $\Vl$ be the \wt/ subspace:
$$
\Vl=\bigl\{v\in V\ |\ E_{ii}\>v=
\bigl(\la_{i-1}-\la_i+\tsum_{m=1}^n\La_i(m)\bigr)\,v\,,\ \,i=1\lc N+1\bigr\}
$$
where $\la_0=\la_{N+1}=0$.
The \KZo/s preserve the \wtd/ of a \gm/ $V$.
Now we are interested in \sol/s to system \(qkz) with values in $\Vl$.
Set $\ell=\sum_{i=1}^N\la_i$. Let
$t=(t_{11}\lc t_{1\la_1},t_{21}\lc t_{2\la_2}\lc t_{N1}\lc t_{N\la_N})\in\Cl$.
The \phf/ is given as follows:
$$
\align
&\Phi(t,z;p) =
\prod^n_{m=1}\,\prod^{N+1}_{i=1}\,\exp\bigl(z_m\mu_i\La_i(m)/p\bigr)\,
\prod^{N}_{i=1}\,\prod^{\la_i}_{j=1}\,
\exp\bigl(t_{ij}(\mu_{i+1}-\mu_i)/p\bigr)\,\,\x
\\
&\x\prod^n_{m=1}\,\prod^{N}_{i=1}\,\prod^{\la_i}_{j=1}
\ {\Gm((t_{ij}-z_m+\La_i(m))/p)\over\Gm((t_{ij}-z_m+\La_{i+1}(m))/p)}\ \x
\\
&\x\,\prod^{N}_{i=1}\,\prod^{\la_i}_{j=2}\,\prod^{j-1}_{k=1}
\ {\Gm((t_{ik}-t_{ij}-1)/p)\over\Gm((t_{ik}-t_{ij}+1)/p)}
\ \prod^{N-1}_{i=1}\,\prod^{\la_i}_{j=1}\,\prod^{\la_{i+1}}_{k=1}
\ {\Gm((t_{i+1,k}-t_{ij}+1)/p)\over\Gm((t_{i+1,k}-t_{ij})/p)}\,.\kern1em
\Tagg{Phi}
\endalign
$$
The \wt/ \fn/ $w(t,z)$ is given by an algebraic construction taken from
the nested \Ba/. In particular, $w(t,z)$ does not depend on $p,\mu$ at all.
 For more details cf\. {\=$^{\Ci{TV},\,\Ci{TV2}}$}.

\Sect Asymptotic \sol/s to \qKZ/
\par\nt
Let $p\to 0$. We are interested in \asol/s to system \(qkz) which have the form
$$
\Psi(z;p)=\exp\bigl(\tau(z)/p\bigr)\tsum_{s=0}^\8\Psi_s(z)\>p^s\,.
\Tag{asol}
$$
The \phf/ has an \asex/:
$$
\Phi(t,z;p)\simeq a(p)\,\exp\bigl(\tau(t,z)/p\bigr)
\,\Xi(t,z)\,\bigl(1+\tsum_{s=1}^\8\pho_s(t,z)\>p^s\bigr)
$$
where $a(p),\,\tau(t,z),\,\Xi(t,z),\,\pho_s(t,z)$ are suitable \fn/s.
\par
A point $(t,z)$ is called a {\it \cp/} if
$\exp\Bigl(\dsize{\der\tau\over\der t_a}(t,z)\Bigr)=1$ for $a=1\lc\ell$.
Set $H(t,z)=\det\Bigl(\dsize{\der^2\tau\over\der t_a\der t_b}(t,z)\Bigr)$.
A \cp/ $(t,z)$ is called {\it \ndg/} if $H(t,z)\ne 0$.
Equations for \cp/s coincide with the \Bae/s in the nested \Ba/.
The set of \cp/s is preserved by the natural action of the product of the
\sym/ groups $\S=\S_{\la_1}\lx\S_{\la_N}$ on variables $t$.
\par
Let $(\ts,\zs)$ be a \ncp/. Set
$\dsize I_a={1\over 2\pi i}\>{\der\tau\over\der t_a}(\ts,\zs)$,
$I(t)=\exp\bigl(-2\pi i\tsum_{a=1}^\ell I_at_a/p\bigr)$ and
$\thu(t,z) = \tau(t,z)-2\pi i\tsum_{a=1}^\ell I_at_a$. Let $D$ be a suitable
small real disk containing $(\ts,\zs)$, $\dim^{\vp1}_\R D=\ell$. Set
$$
\Psi(z;p) = {1\over a(p)}\>\Bigl(-{1\over 2\pi p}\Bigr)^{\ell/2}
\int_DI(t)\>\Phi(t,z;p)\>w(t,z;p)\dt\,.
$$
As $p\to 0$, by the \msd/ $\Psi(z;p)$ has an \asex/
$$
\Psi(z;p) \simeq
\exp\bigl(\thu(t(z),z)/p\bigr)\,\Xi(t(z),z)\,H^{-{1\over2}}(t(z),z)\,
\bigl(w(t(z),z)+\tsum_{s=1}^\8\psi_s(t(z),z)\>p^s\bigr)
$$
where a \fn/ $t(z)$ is \st/ $(t(z),z)$ is a \ncp/ and $t(\zs)=\ts$.
\proclaim{\Th{sol}}
Let $\Phi(t,z;p)w(t,z;p)$ be an \inrp/ for \sol/s to \qKZ/ \(qkz).
The \asex/ of $\Psi(z;p)$ as $p\to 0$ gives an \asol/ to system \(qkz)
of the form \(asol).
\endproclaim
\proclaim{\Lm*} Let $(t,z)$ be a \ncp/. Then
$$
K_i(z;0)w(t,z)=\exp\Bigl({\der\tau\over\der z_i}(t,z)\Bigr)\,
w(t,z)\,,\qquad i=1\lc n\,.
$$
\endproclaim
A \cp/ $(t,z)$ is called an {\it \off/} \cp/ if $t_{ij}\ne t_{ik}$ for
$(i,j)\ne(i,k)$, and a {\it diagonal} \cp/, otherwise.
\proclaim{\Th{zero}}
Let $(\ts,\zs)$ be a diagonal \ncp/. Then\nl
\hbox{ }$\exp\bigl(-\thu(t(z),z)/p\bigr)\>\Psi(z;p)=O(p^\8)$ as $p\to 0$.
\endproclaim
Let $S_i$ be the Shapovalov form on $V_i$. Set $S=S_1\lox S_n$.
Let $\Ka_i(z;p)$ be the dual to $K_i(z;p)$ \wrt/ the form $S$.
Set $\Rb(z)=\prod_{j=2}^n\prod_{i=1}^{\vp n\smash{j-1}}R_{ij}(z_i-z_j)$
both indices in the ordered product increasing from the left to the right.
\proclaim{\Lm{Sh}}
{}\ {\rm i)} $\Ka_i(z;p)=\Rb(z)\,Z_i\bigl(K_i(z;-p)\Rb\1(z)\bigr)$.\nl
{\rm\hbox{\enspace}ii)} \,Operators $\,R_{ij}(z)$ and $\Rb(z)$ are symmetric
\wrt/ the form $S$.
\endproclaim
\nt
Set $\bra w_1,w_2\ket=S(\Rb w_1,w_2)$. Let $(t(z,\mu),z)$ be an \off/ \ncp/.
\proclaim{\Th{norm}}
$\bra w(t(z,\mu),z),w(t(z,\mu),z)\ket=
\const\ \Xi^{-2}(t(z,\mu),z)\,H(t(z,\mu),z)$ where $\const$
does not depend on continuous deformations of the \cp/ $(t(z,\mu),z)$.
\endproclaim
\proclaim{\Cj} For any \off/ \cp/ $(t,z)$
$$
\bra w(t,z),w(t,z)\ket=(-1)^\ell\,\Xi^{-2}(t,z)\,H(t,z)\,.
$$
 For any \cp/s $(t,z)$,\alb$(\tb,z)$ lying in different {\=$\S$-orbits}
$\bra w(\tb,z),w(t,z)\ket=0$.
\endproclaim
\nt
This Conjecture was proved for the $\gg$ case in\,\Cite{TV2} using
the limit $\exp(\mu_2-\mu_1)\to 0$. A combinatorial proof for the first part of
Conjecture for the $\gg$ case was given in\,\Cite{K}, and
for the $\frak{gl}_3$ case (with a special choice of \gm/s) in\,\Cite{R}.
 For similar results for the \difl/ \KZ/ \eq/ cf\&$^{\Ci{RV},\,\Ci{V2}}\!$.
\par
Let $\CCm$ be the set of all \off/ \cp/s modulo the
action of the group $\S$. Vectors $w(t,z)$ are preserved by the action of $\S$
modulo multiplication by a nonzero scalar factor.
\proclaim{\Th{Bethe}}
Let $\g=\gg$. Let $z,\,\mu,\,\La(1)\lc\La(n)$ be generic. Then\nl
$\{w(t,z)\}_{t\in\CCm}$ is a base in $\Vl$.
\endproclaim

\Sect \qKZ/ and bases of singular vectors.
\par\nt
Assume that $\mu=0$. Set
$\sing=\bigl\{v\in V\ |\ E_{i,i+1}\>v=0\,,\ \,i=1\lc N\>\bigr\}$
and $\singl=\Vl\>\cap\>\sing$.
\proclaim{\Lm{sing0}}
%\kern-.6em$^{\Ci{FT2},\,\Ci{KiR},\,\Ci{KR},\,\Ci{Kid}}$\space
Let $(t,z)$ be an \off/ \cp/. Then $w(t,z)\in\singl$.
\endproclaim
Let $\g=\gg$. Let $\CC(z)$ be the set of all \off/ \cp/s modulo the action
of the \sym/ group $\S_\la$. Let $\La_1(m)-\La_2(m)<0$, \,$m=1\lc n$, or
let $\La(1)\lc\La(n)$ be generic.
\proclaim{\Th{Bethe-}}
 For generic $z$ all \off/ \cp/s are \ndg/. \nl
Moreover, $\#\CC(z)=\dims$ and $\{w(t,z)\}_{t\in\CC(z)}$ is a base in $\singl$.
\endproclaim
\nt
Assume that $\La_1(m)-\La_2(m)\in\Z_{\ge0}$, \,$m=1\lc n$.
Let $V_1\lc V_n$ be the \ir/ \gm/ with \hw/s $\La(1)\lc\La(n)$, \resp/.
A \cp/ $(t,z)$ is called a {\it trivial} \cp/ if $w(t,z)=0$,
and a {\it nontrivial} \cp/, otherwise.
Let $\CC(z)$ be the set of all nontrivial \cp/s modulo the action of
the \sym/ group $\S_\la$.
\proclaim{\Th{Bethe+}}
 For any $z$ all \off/ trivial \cp/s are degenerate.\nl
 For generic $z$ all nontrivial \cp/s are \ndg/. Moreover,\nl
$\#\CC(z)=\dims$ and $\{w(t,z)\}_{t\in\CC(z)}$ is a base in $\singl$.
\endproclaim
\nt
 For similar results for the \difl/ \KZ/ \eq/ cf\&$^{\Ci{RV},\,\Ci{V2}}$.

\sect Acknowledgements
\par\nt
We thank Department of Mathematics, University of Tokyo for hospitality.
The first author acknowledges a financial support from the Organizing Committee
of the {\=XX$^{\text{th}}$} ICGTMP (Yamada Conference).

\enddocument